\documentclass[a4paper,11pt]{article}
\pdfoutput=1

\usepackage{jheppub1}
\usepackage[T1]{fontenc}
\usepackage{amssymb}
\usepackage{graphicx}
\usepackage{amsmath}
\usepackage{tensor}
\usepackage{hyperref}
\usepackage{epstopdf}
\usepackage{extarrows}
\usepackage{xcolor}
\usepackage{tensor}
\usepackage{subfigure}

\newcommand{\td}{\text{d}}

\def\U {\hat{U}}
\def\C {\mathcal{C}}

\def\pD {\mathcal{D}}

\def\C {\mathcal{C}}

\begin{document}
\title{What kind of ``complexity'' is dual to holographic complexity?}
\author[a]{Run-Qiu Yang,}
\author[b,c]{Yu-Sen An,}
\author[d]{Chao Niu,}
\author[d]{Cheng-Yong Zhang,}
\author[e]{Keun-Young Kim}

\emailAdd{aqiu@tju.edu.cn}
\emailAdd{anyusen@itp.ac.cn}
\emailAdd{chaoniu09@gmail.com}
\emailAdd{zhangcy@email.jnu.edu.cn}
\emailAdd{fortoe@gist.ac.kr}

\affiliation[a]{Center for Joint Quantum Studies and Department of Physics, School of Science, Tianjin University, Yaguan Road 135, Jinnan District, 300350 Tianjin, P.~R.~China}
\affiliation[b]{Institute of Theoretical Physics, Chinese Academy of Science, Beijing 100190, China}
\affiliation[c]{School of physical Science, University of Chinese Academy of Science, Beijing 100049, China}
\affiliation[d]{Department of Physics and Siyuan Laboratory, Jinan University, Guangzhou 510632, China}
\affiliation[e]{ School of Physics and Chemistry, Gwangju Institute of Science and Technology,
Gwangju 61005, Korea
}

\abstract{
It is assumed that the holographic complexities such as the complexity-action (CA) and the complexity-volume (CV) conjecture are dual to complexity in field theory. However, because the definition of the complexity in field theory is still not complete, the confirmation of the holographic duality of the complexity is ambiguous.
To improve this situation, we approach the problem from a different angle.
We first identify minimal and genuin properties that the filed theory dual of the holographic complexity should satisfy without assuming anything from the circuit complexity or the information theory. Based on these properties, we propose a field theory formula dual to the holographic complexity. Our field theory formula implies that the complexity between certain states in two dimensional CFTs is given by the Liouville action, which is compatible with the path-integral complexity.
It gives natural interpretations for both the CA and CV conjectures and identify what their reference states are.  When applied to the thermo-field double states, it also gives consistent results with the holographic results in the CA conjecture: both the divergent term and finite term.
}

\maketitle

\noindent

\section{Introduction}
For the last decade the quantum informational concepts have been actively applied to the gravity theory including black hole physics. For example, ``quantum complexity'' (``complexity'' in short) has been introduced as a tool to investigate the interior of the black hole. A complexity has been well studied in quantum circuits. Simply speaking, the complexity of an {\it operator} is the minimal number of gates(basic building blocks to construct the circuit) to simulate the operator by quantum circuits. The complexity of {\it state} is the minimal number of gates necessary to transform a reference state to a target state.

The concept of the complexity was first introduce to understand the fire-wall proposal of the black hole~\cite{Harlow:2013tf} and the growth of the Einstein-Rosen bridge of the AdS black holes~\cite{Stanford:2014jda,Susskind:2014rva,Susskind:2014rva2}.
The complexity from holographic perspective is called ``holographic complexity'', which deals with the complexity of  a quantum state dual to the boundary time slice of an eternal asymptotic AdS black hole.
There are two main proposals for holographic complexity: the complexity-volume (CV) conjecture~\cite{Stanford:2014jda} and the complexity-action (CA) conjecture~\cite{Brown:2015bva}.

Let us denote by $t_R$ and $t_L$ the time slices at the right and left boundaries of an asymptotically AdS black hole. In the CV conjecture, the complexity is proportional to the maximum volume of space-like hypersurfaces.    The CV conjecture says
\begin{equation}\label{CVeq1}
  \C=\max_{\partial\Sigma=t_L\cup t_R}\frac{\text{Vol}(\Sigma)}{G_N\ell}\,,
\end{equation}
where $\Sigma$ is a spacelike surface connecting $t_L$ and $t_R$, $G_N$ is the Newton's constant, and $\ell$ is a certain length scale. In the CA conjecture, the complexity is the on-shell action
\begin{equation}\label{CAeq1}
  \C=\frac{S_{\text{WdW,on-shell}}}{\pi\hbar}\,,
\end{equation}
in the Wheeler-DeWitt (WdW) patch, where the WdW patch is the closure of all spacelike surfaces connecting $t_L$ and $t_R$.

There have been many research investigating the conjectures~\eqref{CVeq1} and \eqref{CAeq1}: the structure of the UV divergence of the holographic complexities~\cite{Chapman:2016hwi,Kim:2017lrw,Akhavan:2019zax,Omidi:2020oit}, the time-evolution~\cite{Carmi:2017jqz,Kim:2017qrq,An:2018dbz}, the growth rate of the complexities and the Lloyd's bound~\cite{Cai:2016xho,Yang:2016awy,Pan:2016ecg,Alishahiha:2017hwg,An:2018xhv,Jiang:2018pfk,Jiang:2018sqj, Yang:2019gce, HosseiniMansoori:2018gdu},  the quench effects~\cite{Moosa:2017yvt,Chen:2018mcc,Fan:2018xwf}, the applications to cosmological evolutions~\cite{An:2019opz,Lehners:2020pem}, the behavior in $T\bar{T}$ deformation~\cite{Geng:2019yxo}, the generalization of time-dependent background~\cite{Momeni:2016ira} and applications in cosmology~\cite{An:2019opz}. For other holographic complexity conjectures apart from the CV and CA conjecture see, for example, Refs~\cite{Couch:2016exn,Caputa:2017urj,Caputa:2017yrh,Fan:2018wnv,Fan:2019mbp,Momeni:2016ira}. For  the generalizations of subregion complexities see, for example, ~\cite{Alishahiha:2015rta,Carmi:2016wjl,Geng:2019yxo,Auzzi:2019mah,Auzzi:2019vyh,Ben-Ami:2016qex}.

\begin{figure}[]
\centering
    \subfigure[Method 1]{\includegraphics[width=5.5cm]{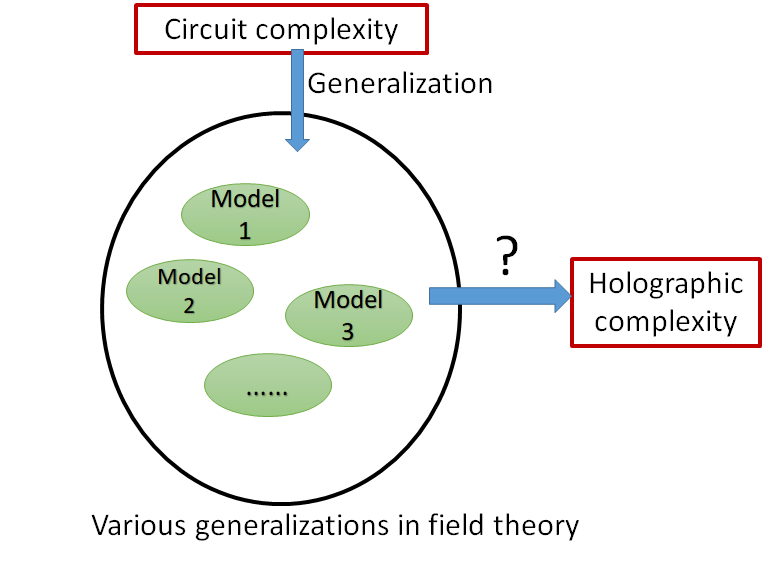}\label{tworoads1}} \ \ \
    \subfigure[Method 2]{\includegraphics[width=5.5cm]{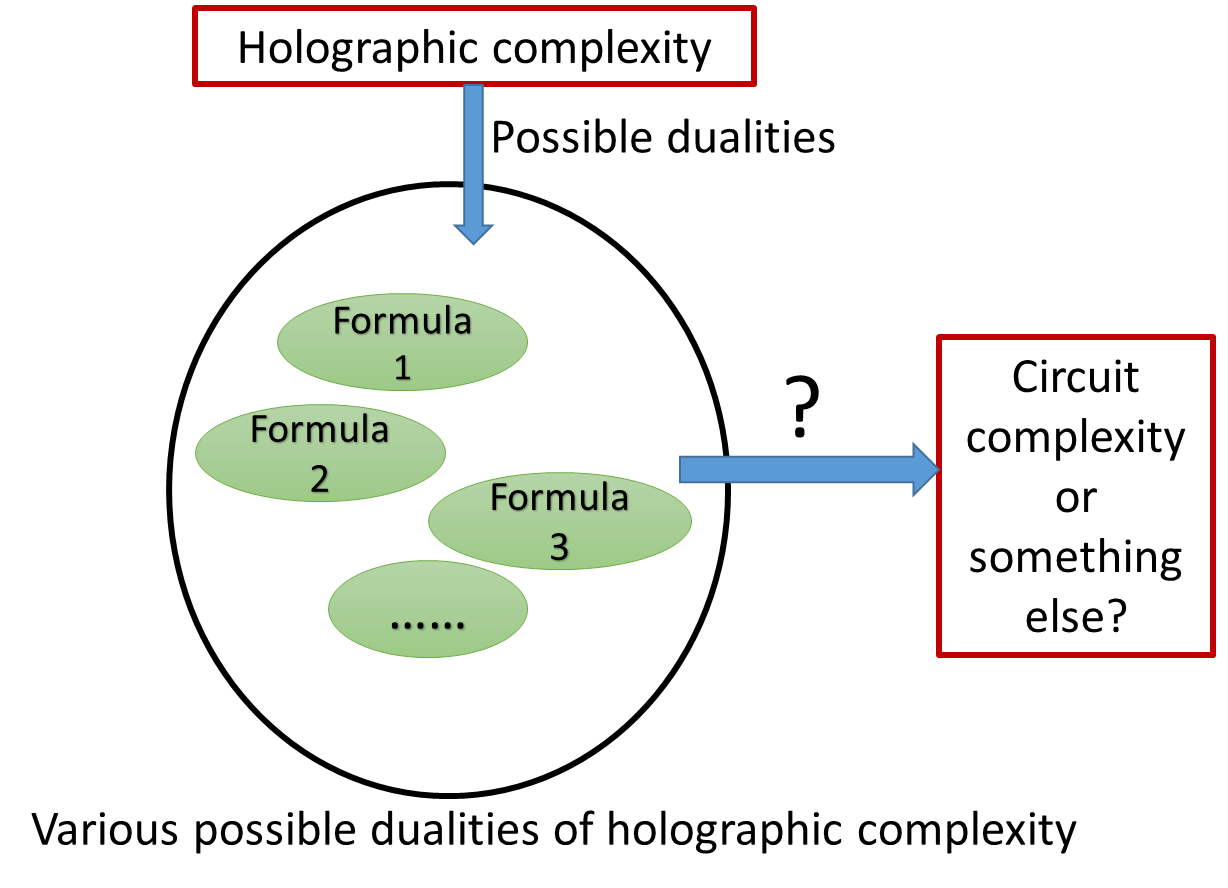}\label{tworoads2}}
   \caption{Two different methods to understand the holographic complexity. (a) Assume that the holographic complexity is a precise dual of a specific generalization of the circuit complexity. We investigate different possible generalizations of the circuit complexity in field theory and then seek for the best one compatible with the holographic complexity. (b) Do not make any assumption on what the field theory dual of the holographic complexity is.  First, just try to find out various possible candidates of the field dual of the holographic complexity. Next, we ask if they have any relationship to the circuit complexity or something else.} \label{tworoads}
\end{figure}

Contrary to various progress on the complexity in gravity side, the complexity theory in quantum field theory (even its precise definition) is still incomplete.  One natural idea to define the complexity in field theory is as follows.
\begin{itemize}
\item[(1)] Start with various well-established models of circuit complexity and generalize them for field theory in certain ways.
\item[(2)] Analyze the consequences of those generalizations and figure out which one is compatible with the holographic complexity.
\end{itemize}
This idea is based on the {\it assumption} that the holographic complexities such as the CV or CA conjectures are indeed dual to a kind of ``circuit complexity''. See Fig.~\ref{tworoads1}.
Following this idea, there have been attempts to generalize the concepts of complexity of discrete quantum circuits to continuous systems:  ``complexity geometry''~\cite{Susskind:2014jwa,Brown:2016wib,Brown:2017jil}, 
Fubini-study metric~\cite{Chapman:2017rqy}, and path-integral optimization~\cite{Caputa:2017urj,Caputa:2017yrh,Bhattacharyya:2018wym,Takayanagi:2018pml}. See also \cite{Hashimoto:2017fga,Hashimoto:2018bmb,Flory:2018akz,Flory:2019kah,Belin:2018fxe,Belin:2018bpg}.

In particular, the complexity geometry is actively investigated. The basic idea was first proposed by Nielsen et al.~\cite{Nielsen1133,Nielsen:2006:GAQ:2011686.2011688,Dowling:2008:GQC:2016985.2016986}, in which the authors considered a continuum approximation of the circuit complexity.  It introduces so-called ``complexity geometry''  and the geodesic distance there. See for examples~\cite{Jefferson:2017sdb,Yang:2017nfn,Reynolds:2017jfs,Kim:2017qrq,Khan:2018rzm,Hackl:2018ptj,Yang:2018nda,Yang:2018tpo,Alves:2018qfv,Magan:2018nmu,Auzzi:2020idm}. Along this road, there have been many works showing positive supports in identifying the field theory complexity in the sense of the agreement with the holographic complexity~\cite{Caputa:2018kdj,Camargo:2018eof,Guo:2018kzl,Bhattacharyya:2018bbv, Jiang:2018gft,Chapman:2018hou,Ali:2018fcz,Chapman:2018dem,Doroudiani:2019llj,Auzzi:2020idm}. However, a few fundamental questions still remain.
%


First, what are the reference states in the holographic conjectures?  For both the CV and CA conjectures, the target state is dual to the thermofield double (TFD) state~\cite{Maldacena:2001kr} but, the reference state is not known. Because these holographic complexities are supposed to be the complexity of states, the reference state needs to be clearly identified.

Second, what makes the differences in the time-evolution between the CV and CA conjectures? In both  cases, the complexity grows linearly in time at late time. However, they behave differently at early time~\cite{Carmi:2017jqz,Kim:2017qrq}, which indicates that two conjectures may correspond to different physical quantities (complexity or something else) in field theory.

There is another important issue to consider when we identify the field theory dual of the holographic complexity: the properties of the proposed complexity in field theory may differ for different models.
The ``path-integral complexity''~\cite{Caputa:2017urj,Caputa:2017yrh,Bhattacharyya:2018wym}, which is different from Nielsen's, is considered to describe the complexity between the ground state and the field operator eigenstate in a two-dimensional conformal field theory (CFT). Here, the ground state is build by the tensor network renormalizations~\cite{PhysRevLett.115.180405} and the complexity is identified with the on-shell Liouville action.
One of the important properties of the  path-integral complexity is that it is \textit{unitary invariant} and \textit{bases-independent}. By minimizing such a complexity, the Einstein's equation in 2+1 dimension can be obtained~\cite{Czech:2017ryf}. For the relation between the path integral complexity and circuit complexity see Ref.~\cite{Camargo:2019isp}.

\begin{figure}[]
\centering
    \subfigure[Possibility 1]{\includegraphics[width=5.5cm]{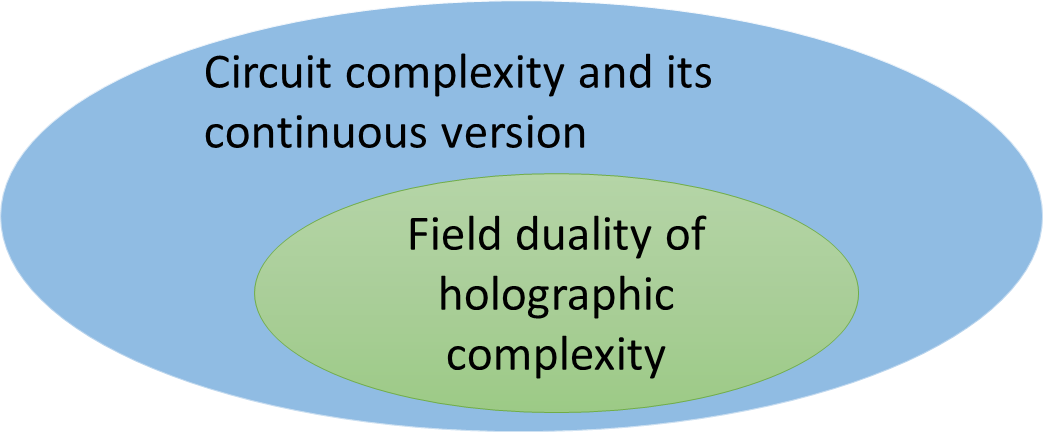}\label{newpossib1}} \ \ \
    \subfigure[Possibility 2]{\includegraphics[width=6.5cm]{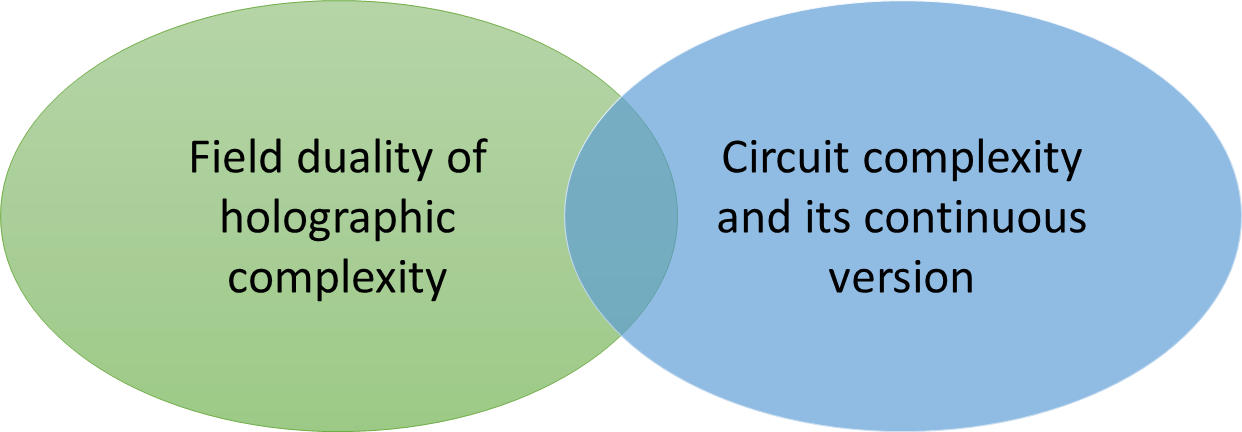}\label{newpossib2}}
  \caption{A schematic explanations for possible relationships between the circuit complexity and the field-theory duals of the holographic complexity. (a) The field theory dual of the holographic complexity belongs to the continuous version of the circuit complexity. (b) The field theory dual of the holographic complexity and a continuous version of the circuit complexity share some common properties, but, there are still differences too.} \label{newpossib}
\end{figure}

Though a large amount of efforts have been made along the road shown in  Fig.~\ref{tworoads1}, there is still no proposal which can be completely compatible with any holographic complexity.
It motivated us to step back and ask a question: is the quantity so-called  the ``holographic complexity'' really a ``complexity'' in field theory? Even if it is the case, it is possible that that the field theory dual of the holographic complexity may belong to a different type of complexity, which is not necessarily the same as a continuous version of the circuit  complexity. In this case, the field theory dual of the holographic complexity and the continuous version of the circuit complexity may share some common properties but there may be differences too.
See Fig.~\ref{newpossib} for a schematic explanation.  For now, it seems that there is no evidence to rule out any one in Fig.~\ref{newpossib}. Thus, let us keep this possibilities open.

The way in  Fig.~\ref{tworoads1} will work only for the
case in Fig.~\ref{newpossib1}. If the relationship shown in Fig.~\ref{newpossib2} is valid, then the way in Fig.~\ref{tworoads1} is not suitable. Instead, a different method shown in Fig.~\ref{tworoads2} is more promising:
\begin{itemize}
\item[(1)] First try to find all possible field theory duals of the holographic complexity, by considering the genuin properties of the holographic complexity,  without assuming anything from field theory.
\item[(2)] Next, we check if there is any candidate which can match with the basic requirements of the circuit complexity (non-negativity, right-invariance, triangle inequality, et.)
\end{itemize}
This method indeed can cover both possibilities in Fig.~\ref{newpossib}.
The main goal of this paper is to make a step towards this new road.

We assume that the holographic complexity ($\C_V$ or $\C_A$) has a field theory dual denoted by $\bar{\C}$
\begin{equation}\label{eqCVAF}
  \C_{V(A)}=\bar{\C}(|\psi\rangle,|R\rangle)\,.
\end{equation}
which describes an unknown relationship between a target state $|\psi\rangle$ and a reference state $|R\rangle$.  In this paper, we use the notation $\bar{\C}$ instead of $\C$, to denote it is the specific `complexity' related with the holographic complexity. In principle, it may be different from the usual circuit complexity or any other quantum computational concept.
Our first task is to find if there is any special properties of this function $\bar{\C}$. We emphasize again that we do not assume anything from field theory. i.e. at this stage $\bar{\C}(|\psi\rangle,|R\rangle)$ may not correspond to a specific kind of ``circuit complexity''.
We just try to investigate the properties of the holographic complexity and find out what the possible candidates of the function $\bar{\C}$ are.

First, we follow a usual way in theoretical physics: \textit{symmetry is  important}. We will argue that, the diffeomorphic invariance of the holographic complexity implies that, at least for a large class of infinite dimensional unitary group (strictly speaking, faithful unitary representation of an infinite dimensional Lie group) $\mathcal{G}$, the function $\bar{\C}$ is invariant under a transformation of $\mathcal{G}$:
\begin{equation}\label{property1}
  \forall \U\in\mathcal{G},~~~\bar{\C}(|\psi\rangle,|R\rangle)=\bar{\C}(\U|\psi\rangle,\U|R\rangle)\,.
\end{equation}
This property of the holographic complexity, however, cannot be read from the circuit complexity. This shows that the field theory dual of the holographic complexity has infinitely many symmetries so infinite constraints.

The second important property of $\bar{\C}$ is the it is an extensive quantity for product states
\begin{equation}\label{property2}
  \bar{\C}(|\psi_1\rangle\otimes|\psi_2\rangle,|R_1\rangle\otimes|R_2\rangle))=\bar{\C}(|\psi_1\rangle,|R_1\rangle)+\bar{\C}(|\psi_2\rangle,|R_2\rangle)\,.
\end{equation}
This property comes from a fact that the holographic complexity is proportional to the volume of the boundary slice. This is also a special property of the holographic complexity and shows an essential difference from the entanglement entropy.

By combining these two basic properties, we propose a class of possible simple candidates for the function $\bar{\C}$ and choose  particular forms as examples to make detailed discussion. Our proposal have many interesting implications. It supports that the complexity in 2D CFTs can be expressed by the Liouville action, consistent with the path-integral complexity.  It also provides natural interpretations of the CV and CA conjectures and  clarifies their reference and target states. i.e. our proposal may answer two aforementioned problems at the bottom of page 2. 


The paper is organized as follows: In Sec.~\ref{C-states}, we explain why the holographic complexity implies the equations~\eqref{property1} and \eqref{property2}.  In Sec.~\ref{barC1} we give a class of simple candidates for the function $\bar{\C}$ and show that they can exhibit very rich contents including the new interpretation of the CV and CA conjecture. In Sec.~\ref{tfds1} we apply our proposal to the TFD states and do further consistency checks with holographic complexities.  We conclude in Sec.~\ref{summ}.

\section{Basic properties of holographic complexity}\label{C-states}
\subsection{Holographic complexity is diffeomorphic invariant}

One standard way to investigate physical systems of which structures are not well known is to start with symmetry. Along this line, we may ask if there is any universal symmetry in the CV and CA conjectures? We think the answer is positive.
%
%
%
In both CV and CA conjectures, the complexity is given as a geometric quantity of the bulk spacetime. This implies that there is an important and universal symmetry in the holographic complexity: it is invariant under a bulk diffeomorphic transformations. If we make a bulk diffeomorphic transformation, the boundary theory will be also transformed by the induced transformation.\footnote{In this paper, when we talk about spacetime transformation, we always use ``active viewpoint'': the coordinates is fixed but metric is transformed. We will also use the ``active viewpoint'' to consider transformations of states in the Hilbert space and generators in Lie algebra: the bases of Hilbert space and Lie algebra are unchanged but the quantum states and generators are changed.}

Let us consider a simple case in a pure AdS$_{d+1}$ spacetime. There are two kinds of transformations: a conformal group in a boundary theory and an isometric group in AdS spacetime. The two groups are both isomorphic to SO(2,$d$). Every transformation of SO$(2,d)$ in the bulk corresponds to  a conformal transformation at the boundary CFT and versa vice.
Though CFT quantities such as correlation functions, generating functional and partition function are invariant, the operators and quantum states will obtain a SO(2,$d$) transformation and, in general, this transformation is not an identical transformation.\footnote{For example, the gauge transformation of  a U(1) gauge theory does not change the partition function, correlation functions but it will induce a unitary transformation in the Hilbert space. The gauge transformation: $\{\vec{A},\phi\}\mapsto\{\vec{A}+\vec{\nabla}\Lambda, \phi-\partial_t\Lambda\}$ will induce a unitary transformation in the Hamiltonian and quantum states: $H\mapsto\U H\U^\dagger$ and $|\psi\rangle\mapsto\U|\psi\rangle$ with $\U=\exp(iq\Lambda)$. Here $q$ is the charge.}

Suppose that the states $|\psi\rangle$ and $|R\rangle$ are a target state and a reference state in a CFT, of which the `complexity' is given by $\bar{\C}(|\psi\rangle,|R\rangle)$. By the holographic duality, in the corresponding bulk theory, there are bulk metric $g_{\mu\nu}$ and matter fields $A_i$ ($i$ stands for different matter fields). By using the CV or CA conjecture $\bar\C(|\psi\rangle,|R\rangle)$ can be computed as
\begin{equation}\label{So2dCVA100}
  \bar{\C}(|\psi\rangle,|R\rangle)=\C_{V(A)}(g_{\mu\nu},A_i)\,,
\end{equation}
where $\C_{V(A)}(g_{\mu\nu}, A_i)$ stands for the holographic complexity computed by the metric $g_{\mu\nu}$ in the CV or CA conjecture.

Suppose now that $\U_\phi$ is an SO$(2,d)$ transformation in the CFT Hilbert space, which transforms $|\psi\rangle$ and $|\psi\rangle$ to $\U_\phi |\psi\rangle$ and $\U_\phi|R\rangle$. Accordingly, there is a corresponding bulk diffeomorphism $\phi$, which induces a pull-back transformation $\phi^*$ for the the bulk metric $g_{\mu\nu}\mapsto\phi^*(g_{\mu\nu})$ and for matter fields $A_i\rightarrow \phi^*(A_i)$. By symmetry, we have
\begin{equation}\label{So2dCVA1}
 \bar{\C}(\U_\phi|\psi\rangle,\U_\phi|R\rangle)=\C_{V(A)}(\phi^* (g_{\mu\nu}),\phi^*(A_i)) \,.
\end{equation}
In the pure AdS case, the diffeomorphism $\phi$ is an isometry $\phi^* (g_{\mu\nu})=g_{\mu\nu}$. Thus, we have the following important result on the field theory dual of the holographic complexity
\begin{equation}\label{So2dCVA1}
  \forall \U\in \text{SO}(2,d),~~~\bar{\C}(|\psi\rangle,|R\rangle)=\bar{\C}(\U|\psi\rangle,\U|R\rangle)\,.
\end{equation}
%


The same result holds even if the diffeomorphism is not an isometry. Suppose that $\mathcal{M}_{d+1}$ is an arbitrary asymptotically AdS spacetime and $\phi:\mathcal{M}_{d+1}\mapsto\mathcal{M}_{d+1}$ is an arbitrary diffeomorphic transformation, which transforms the boundary time slices and bulk metric $\{t_L,t_R,g_{\mu\nu}\}$ into $\{\phi (t_L),\phi(t_R),\phi^*(g_{\mu\nu})\}$. The holographic complexity computed by them are the  same, i.e.,
\begin{equation}\label{diffeCVCA}
  \C_{V(A)}\{t_L,t_R,g_{\mu\nu}\}=\C_{V(A)}\{\phi (t_L),\phi(t_R),\phi^*(g_{\mu\nu})\}\,.
\end{equation}
This transformation will also induce a transformation on the Hilbert space in the boundary theory, i.e., $\U_\phi:\mathcal{H}\mapsto\mathcal{H}$. This transformation must be unitary as no information will be lost by the diffeomorphic transformation. If $\{|\psi\rangle,|R\rangle\}\subset\mathcal{H}$ is a pair of a target state and a reference state, the diffeomorphic transformation $\phi$ will induce a new pair of target state and reference state $\{\U_\phi|\psi\rangle,\U_\phi|R\rangle\}\subset\mathcal{H}$. Then, we have the following equations for the function $\bar{\C}$
\begin{equation}\label{diffeCVCA2}
  \bar{\C}(|\psi\rangle,|R\rangle)=\C_{V(A)}\{t_L,t_R,g_{\mu\nu}\},~~\bar{\C}(\U_\phi|\psi\rangle,\U_\phi|R\rangle)=\C_{V(A)}\{\phi (t_L),\phi(t_R),\phi^*(g_{\mu\nu})\}\,,
\end{equation}
which imply
\begin{equation}
\bar{\C}(|\psi\rangle,|R\rangle)=\bar{\C}(\U_\phi|\psi\rangle,\U_\phi|R\rangle)\,.
\end{equation}
%
%
All bulk diffeomorphic transformations form an infinite dimensional Lie group, which induces an infinite dimensional unitary group (strictly speaking, a faithful unitary representation) $\mathcal{G}$ on the boundary Hilbert space. Thus, we have the following symmetry for the field theory dual of the holographic complexity
\begin{equation}\label{diffeCVCA3}
 \forall\U\in\mathcal{G},~~~~ \bar{\C}(|\psi\rangle,|R\rangle)=\bar{\C}(\U|\psi\rangle,\U|R\rangle)\,.
\end{equation}

Note that the holographic complexity gives a strong condition to its field theory dual: it must have infinitely many constraints from the symmetries, which may imply the following. If the field theory dual of the holographic complexity is a kind of continuous version of the usual circuit complexity, these constraints may not allow us to choose the gates and penalties in the circuit complexity artificially. From the perspective of the usual circuit complexity, it seems too strong because, in the usual circuit, the gates and penalties can be arbitrarily chosen and, in general, the complexity is not invariant under unitary transformations. However, again, our strategy here is not to have any prejudice from the field theoretic or quantum computational concept. We want to figure out where the holographic complexity leads us, wherever it is.


\subsection{Holographic complexity is extensive}
Another important property we learn from the holographic complexity is that the complexity between an unknown reference state and a boundary state  is proportional to the volume at the boundary time slices, if the volume is large enough and the boundary state is uniform, i.e.,
\begin{equation}\label{CproptoV1}
  \bar{\C}(|\psi\rangle,|R\rangle)\propto V_{\text{bd}},~~~\text{if}~V_{\text{bd}}\rightarrow\infty\,,
\end{equation}
where, $V_{\text{bd}}$ is the {volume of boundary slices}, {\it not} the volume in any bulk region. This property shows an important difference compared with the entanglement entropy, as the latter in general is proportional to the area. Since the volume is an extensive quantity of physical systems, this implies that holographic complexity may be also an extensive quantity. We can prove this statement holographically as follows.

\begin{figure}
  \centering
  \includegraphics[width=.6\textwidth]{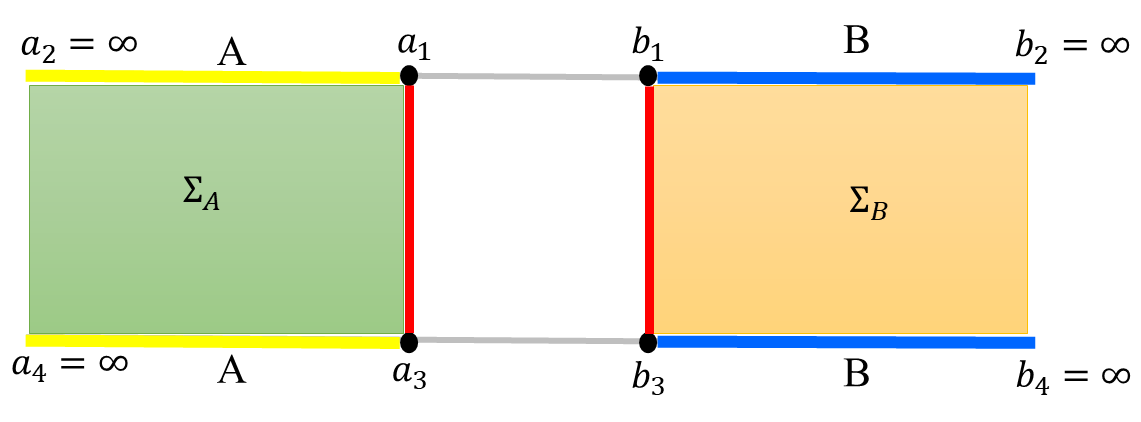}
  \caption{A two-side static black hole which has two disconnected subregions in every boundary. The lines $a_4b_4$ and $a_2b_2$ stand for the left and right AdS boundaries and $A$ and $B$ stand for two subregions of boundary slices. We assume that the boundary slices lay on $t = 0$ hypersurface. $\Sigma_A$ and $\Sigma_B$ are extremal co-dimensional-1 surfaces. $a_1a_3$ and $b_1b_3$ are two RT surfaces.} \label{subAB2}
\end{figure}

Let us consider the complexity between two states  $|R\rangle$  and  $|\psi\rangle$ that contain two independent sub-systems $A$ and $B$.
The systems $A$ and $B$ are locally the same and have the volume $V_A$ and $V_B$, respectively. When two sub-systems are separated far enough, the target state ($|\psi\rangle$) and the reference state ($|R\rangle$) of $A\cup B$ can be written in terms of the direct product of two independent sub-systems:
\begin{equation}\label{subdensity1}
  |\psi\rangle=|\psi\rangle_A\otimes|\psi\rangle_B\,, \qquad |R\rangle=|R\rangle_A\otimes|R\rangle_B\,.
\end{equation}
To deal with the case that a system contains two subregions, we refer to the holographic proposals of the subregion complexities. They have been studied in detail by
Refs.~\cite{Alishahiha:2015rta,Carmi:2016wjl,Geng:2019yxo,Auzzi:2019mah,Auzzi:2019vyh,Ben-Ami:2016qex}. Here, we use the subregion complexity of the CV conjecture as an example. Let us consider a two-side black hole shown in Fig.~\ref{subAB2}, of which the two boundaries are labeled by the lines $a_2b_2$ and $a_4b_4$. The boundary CFT lies on $A\cup B$.
For a static bulk geometry, the CV conjecture for subregions evaluates the volume of the extremal co-dimensional-1 surface in the bulk which is bounded by the subregion on the asymptotic boundary and
the Ryu-Takayanagi (RT) surface for this subregion~\cite{Alishahiha:2015rta}, i.e. for a subregion $A$ at the boundary slice, the subregion complexity by the CV conjecture reads
\begin{equation}\label{subcvs1}
  \bar{\C}(A)\propto\max_{\partial\Sigma=\mathrm{RT(A)}\cup A}\mathrm{Volum}(\Sigma)\,,
\end{equation}
where RT$(A)$ is the RT surface of the subregion $A$. When the two subregions are separated far enough, the RT surfaces will become the type shown by the red lines of Fig.~\ref{subAB2}. This corresponds to the fact that the boundary slices describe a product state. It is clear the total co-dimensional-1 surface contains two disconnected co-dimensional-1 surfaces and
\begin{equation}\label{cvads1}
  \bar{\C}(A\cup B)=\bar{\C}(A)+\bar{\C}(B)\,,
\end{equation}
where $\bar{\C}(A)$ is given by the extremal volume of a co-dimensional-one surface $\Sigma_A$ which is bounded by $a_2a_1a_3a_4$ and $\bar{\C}(B)$ is given by the extremal volume of a co-dimensional-one surface $\Sigma_B$ which is bounded by $b_2b_1b_3b_4$.
The same result can also be obtained by using the subregion CA conjecture~\cite{Carmi:2016wjl}. Thus we have%
%
%
\begin{enumerate}
\item [] {\bf Extensive property}: the complexity of the product states of continuous systems is extensive
    i.e.,
\begin{equation}\label{relCVABs2b}
  \bar{\C}(|\psi\rangle_A\otimes|\psi\rangle_B, |R\rangle_A\otimes|R\rangle_B)=\bar{\C}(|\psi\rangle_A,|R\rangle_A)+\bar{\C}(|\psi\rangle_B,|R\rangle_B)\,,
\end{equation}
%
\end{enumerate}
%
%
This property is not easy to be found if we think only from the perspective of circuits complexity.
If the field theory dual of the holographic complexity is a kind of continuous version of the circuit complexity, it should have very special properties which do not usually (easily) appear in the studies of the circuit complexity.

\section{Field theory dual of the holographic complexity and applications}\label{barC1}
\subsection{Proposal and infinitesimal triangle inequality}\label{proptrig}
We have clarified two basic properties of a field theory dual of the holographic complexity: Eq.~\eqref{diffeCVCA3} and Eq.~\eqref{relCVABs2b}. The next task is to find possible mathematical formulas for $\bar{\C}$ satisfying those properties.  Since Eq.~\eqref{diffeCVCA3} and Eq.~\eqref{relCVABs2b} do not lead to a unique formula for $\bar{\C}$ we want to propose one possibility which looks minimal and most relevant to our purpose.

Note that the unitary group $\mathcal{G}$, which is a faithful unitary representation of diffeomorphism group of $\mathcal{M}_{d+1}$, is a very large group. If we assume that a field theory dual of the holographic complexity should have a natural mathematical form, then a simple choice for  $\bar{\C}$ is
\begin{equation}\label{onefC1}
  \bar{\C}(|\psi\rangle,|R\rangle)=f(\langle\psi|R\rangle)\,,
\end{equation}
where $f$ is an unknown function. If we combine it with the property \eqref{relCVABs2b}, we conclude that for a complex number $x$, the simplest $f(x)$ is
\begin{equation}
f(x) = \alpha_1\text{Re}\ln x+\alpha_2|\text{Im}\ln x| \,,
\end{equation}
for any constant real numbers $\alpha_1$ and $\alpha_2$. By choosing $\alpha_1 =-1, \alpha_2=1$ we have
\begin{equation}\label{defholoCeq2}
  \bar{\C}(|\psi\rangle,|R\rangle)=-\text{Re}\ln x+|\text{Im}\ln x|=- \ln|\langle\psi|R\rangle|+|\text{Im}\ln\langle\psi|R\rangle|\,.
\end{equation}
As the complexity has the freedom of an overall factor, only the ratio $\alpha_1/\alpha_2$ is relevant. We assume $\alpha_1/\alpha_2=-1$  in this paper.

\begin{figure}
  \centering
  \includegraphics[width=0.4\textwidth]{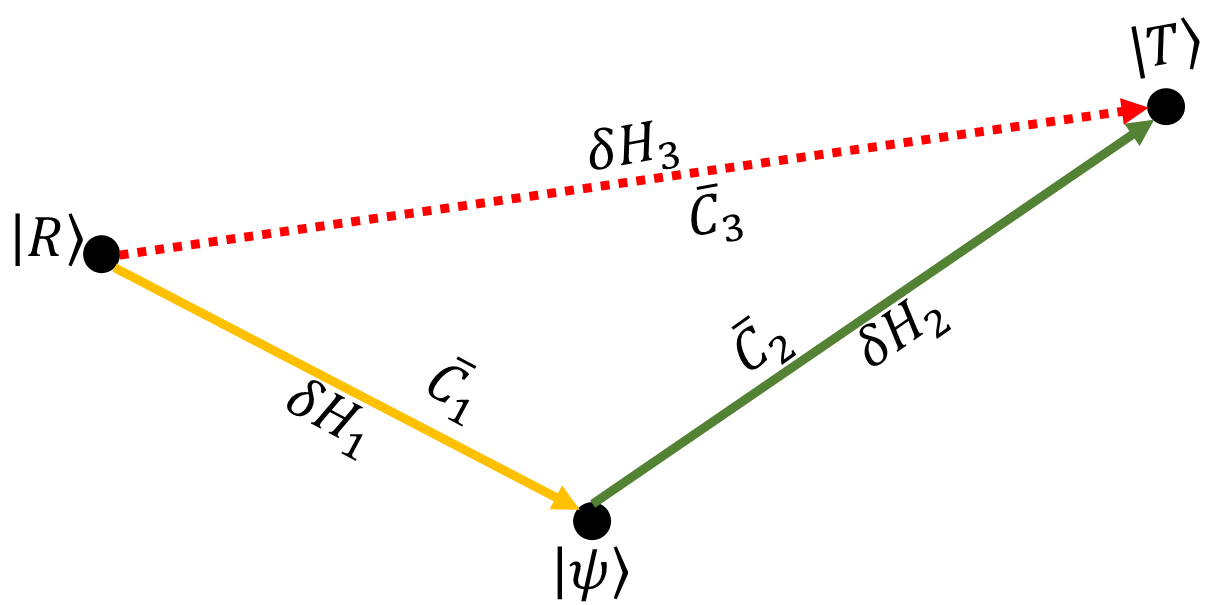}
  \caption{Schematic explanation about the proof of triangle inequality } \label{trig}
\end{figure}

Usually, ``complexity'' stands for a kind of ``distance'' so is expected to satisfy the triangle inequality.
%
Indeed, Eq.~\eqref{defholoCeq2} satisfies the triangle inequality under certain conditions.
Let us first prove the ``infinitesimal version'' of the triangle inequality.
Let us consider arbitrary three infinitesimally close quantum states $|R\rangle, |T\rangle$ and $|\psi\rangle$ as shown in Fig.~\ref{trig}.  There may be  three hermitian Hamiltonians (may not be unique) $\{H_1, H_2, H_3\}$ and an infinitesimal parameter $\delta>0$, which satisfy
\begin{equation}\label{relh3s1}
  |\psi\rangle=e^{-i\delta H_1}|R\rangle,~~|T\rangle=e^{-i\delta H_2}|\psi\rangle=e^{-i\delta H_3}|R\rangle\,.
\end{equation}
The complexities between these three quantum states are labeled by $\bar{\C}_1, \bar{\C}_2$ and $\bar{\C}_3$. See Fig.~\ref{trig}.

Because
\begin{equation}\label{innerrpsi1}
  \langle\psi|R\rangle=\langle R|e^{i\delta H_1}|R\rangle=1+i\delta\langle R|H_1|R\rangle+\mathcal{O}(\delta^2) \,,
\end{equation}
our proposal \eqref{defholoCeq2} gives
\begin{equation}\label{cthrees1}
  \bar{\C}_1=\delta\left|\langle R|H_1|R\rangle\right|+\mathcal{O}(\delta^2)\,.
\end{equation}
Similarly, we have
\begin{equation}\label{cthrees2}
  \bar{\C}_2=\delta\left|\langle \psi|H_2|\psi\rangle\right|+\mathcal{O}(\delta^2)\,, \qquad \bar{\C}_3=\delta\left|\langle R|H_3|R\rangle\right|+\mathcal{O}(\delta^2)\,.
\end{equation}
$\bar{\C}_2$ can be also expressed as
\begin{equation}\label{cthrees2}
  \bar{\C}_2=\delta\left|\langle R|H_2|R\rangle\right|+\mathcal{O}(\delta^2)\,,
\end{equation}
because
\begin{equation}
\langle \psi|H_2|\psi\rangle=\langle R|e^{i\delta H_1}H_2e^{-i\delta H_1}|R\rangle=\langle R|H_2|R\rangle+\mathcal{O}(\delta) \,.
\end{equation}
Furthermore, $ \bar{\C}_3$ can be expressed as
\begin{equation}\label{cthrees2b}
  \bar{\C}_3=\delta\left|\langle R|H_2+H_1|R\rangle\right|+\mathcal{O}(\delta^2)\,.
\end{equation}
because, up to order $\mathcal{O}(\delta)$,
\begin{equation}\label{h1h2h31}
  \langle R|T\rangle=1-i\delta\langle R|H_3|R\rangle=\langle R|e^{-i\delta H_2} e^{-i\delta H_1}|R\rangle=1-i\delta\langle R|H_2+H_1|R\rangle\,,
\end{equation}
which yields $\langle R|H_3|R\rangle=\langle R|H_2+H_1|R\rangle$.
Using the fact
\begin{equation}\label{abstrig}
  \left|\langle R|H_1|R\rangle\right|+\left|\langle R|H_2|R\rangle\right|\geq \left|\langle R|H_2|R\rangle+\langle R|H_2|R\rangle\right|=\left|\langle R|H_1+H_2|R\rangle\right|\,,
\end{equation}
we have
\begin{equation}\label{trigineqc1}
  \bar{\C}_1+\bar{\C}_2\geq \bar{\C}_3\,.
\end{equation}
This shows that our proposal satisfies the triangle inequality for infinitesimally close states. Note that this ``infinitesimal triangle inequality'' does not imply the triangle inequality for arbitrary states. The proof of general triangle inequality needs more preparations and we will come back to this point at end of Sec.~\ref{trigineq}.

\paragraph{A few comments} Let us make a few comments on  Eq.~\eqref{defholoCeq2}. From the perspective of the usual circuit complexity, Eq.~\eqref{defholoCeq2} is too simple and may lose many interesting properties of the circuit complexity.
This is true. However, we recall that the motivation of this paper: we do not study how to use the properties of the usual circuit complexity to recover the holographic results. Instead, we try to understand what a possible field theory dual of the holographic complexity is, whatever it is. Even if it turns out to be  a kind of circuit complexity, it must be a very ``special'' circuit complexity because it contains  properties such as \eqref{diffeCVCA3} and \eqref{relCVABs2b}  which do not appear in the usual circuit complexity.

In Eq.~\eqref{defholoCeq2}  there is an imaginary part of a complex number.
If the inner product $\langle\psi|R\rangle$ is not a real number, there may be ambiguities in two aspects. The first one is due to the multiple valued function ``$\ln(\cdot)$''. For any complex number $x=\rho e^{i\theta}$, we have $\ln x=\ln\rho+i\theta+2n\pi i$ with $n=0,\pm1,\pm2,\cdots$. The second one is due to the fact that two state vectors $|\psi\rangle$ and $e^{i\theta}|\psi\rangle$ describe the same physical state.
These two aspects imply that our formula~\eqref{defholoCeq2} has an ambiguity of adding arbitrary constants.

Interestingly, this ambiguity may correspond to a fact  that, in the CA conjecture, the action of theory has a freedom of adding arbitrary constants.
The CA conjecture connects the on-shell action to the complexity and the action has a freedom of adding a constant term. Thus, any complexity theory, if it is claimed to be dual to the CA conjecture, must have a freedom of adding arbitrary constant.
In the Nielsen's complexity geometry, we  has a freedom to choose an overall factor but do not have a freedom to add a constant. From this perspective,
it seems difficult to use  the Nielsen's ``complexity geometry'' to find a dual of the CA conjecture.

If we consider a continuous ``time'' dependent state $|\psi(t)\rangle$ with $|\psi(0)\rangle=|R\rangle$, then the complexity $\bar{\C}(t)$ will depend on $t$, too. It is natural to require $\bar{\C}(t)$ is also the continuous function and $\bar{\C}(0)=0$. In this case, the ambiguity of a phase factor disappears.

In the following sections, we will show that  Eq.~\eqref{defholoCeq2} indeed can be understood as a kind of ``complexity'', i.e., a kind of minimal ``cost''. However, in general it will not be a usual circuit complexity.
We will investigate some implications of the proposal \eqref{defholoCeq2}, which support that the proposal \eqref{defholoCeq2} may be a correct field theory dual of the  holographic complexity.



\subsection{Path-integral formula and triangle inequality}\label{trigineq}
Here, we compute the complexity by using the path integral formulation.
%
Let us take a normalized initial state $|\psi_0\rangle$, a target state $|\psi(t)\rangle$ and a time evolution operator $\hat{U}(t)$.
We may consider a one dimensional quantum mechanical system without loss of generality and the Feynman propagator yields
\begin{equation}\label{pathforK1}
  K(x_2,t_2;x_1,0):=\langle x_2|\hat{U}(t)|x_1\rangle=\frac1{\mathcal{N}}\int_{x(0)=x_1}^{x(t_2)=x_2}\pD[x]\exp\left\{\frac{i}{\hbar}S[x(t)]\right\} \,,
\end{equation}
where  $\mathcal{N}$ is a normalization factor and $S[x(t)]$ is the classical action functional.  The transition amplitude reads  
\begin{equation}
Z := \langle \psi(t) | \psi_0 \rangle  = \iint\td x_2\td x_2\psi_0^*(x_2)K(x_2,t_2;x_1,0)\psi_0(x_1)\,,
\end{equation}
where $\psi_0(x):=\langle x|\psi_0\rangle$ denotes the wave function of the initial state. By Eq.~\eqref{defholoCeq2}, the complexity between $|\psi_0\rangle$ and $|\psi(t)\rangle$ is
\begin{equation}\label{waveCm1}
  \bar{\C}(t_2)=-\ln|Z|+|\text{Im}\ln Z|\,,
\end{equation}
so the time-dependent complexity is determined by the the action and the initial state.

In quantum field theory, a similar procedure works. The complexity between the states $|\Psi\rangle$ and $|\Phi\rangle$ can be expressed in terms of a functional integration
\begin{equation}\label{modCforscalar}
  \bar{\C}=-\ln|Z|+\text{Im}\ln Z\,,
\end{equation}
and
\begin{equation}
{Z = \langle \Phi | \Psi \rangle }= \int\pD[\varphi(x)]\Phi^*[\varphi(x)]\Psi[\varphi(x)]
\end{equation}
where $\Psi[\varphi(x)]=\langle\varphi|\Psi\rangle$ is the wave functional of $|\Psi\rangle$. The complexity between the time dependent state $|\Psi(t)\rangle$ and $|\Psi_0\rangle=|\Psi(0)\rangle$  is Eq.~\eqref{modCforscalar} with
\begin{equation}\label{CmPsi1}
 { Z = \langle \Psi(t) | \Psi(0) \rangle } =\int\pD[\varphi_1(x)]\pD[\varphi_2(x)]\Psi_0^*[\varphi_2(x)]\Psi_0[\varphi_1(x)]K[\varphi_2(x),t_2;\varphi_1(x),0]\,,
\end{equation}
where
\begin{equation}\label{pathintphi}
  K[\varphi_2(x),t_2;\varphi_1(x),t_1]=\frac1{\mathcal{N}[\varphi_1,\varphi_2]}\int_{\varphi(x,t_1)=\varphi_1(x)}^{\varphi(x,t_2)=\varphi_2(x)}\pD[\varphi(x)]\exp\left\{\frac{i}{\hbar}S[\varphi]\right\}\,.
\end{equation}
Here, $\mathcal{N}[\varphi_1,\varphi_2]$ is the normalization factor and satisfies $\mathcal{N}[\varphi_1,\varphi_1]=1$. If we consider the classical limit $\hbar\rightarrow0$ and assume the target state and reference state are eigenstates of the field operator, we have the following approximation
\begin{equation}\label{approxmZ1}
  Z=K\approx\frac1{\mathcal{N}[\varphi_1,\varphi_2]}\exp\left\{\frac{i}{\hbar}S_{\text{cl}}[\varphi]\right\}\,,
\end{equation}
where $S_{\text{cl}}[\varphi]$ is the classical on-shell action. Up to the leading order of $\hbar$, we have
\begin{align}\label{Conshell1}
 \bar{\C} &\approx \frac{1}{\hbar}\min\{|S_{\text{cl}}[\varphi]| \   | \ \forall\varphi(x,t), \ s.t. \ \varphi(x,t_1)=\varphi_1(x),~\varphi(x,t_2)=\varphi_2(x)\}\,,
\end{align}
Here we assume Re$(\ln\mathcal{N}[\varphi_1,\varphi_2])\ll S_{\text{cl}}/\hbar$ and Im$(\ln\mathcal{N}[\varphi_1,\varphi_2])\ll S_{\text{cl}}/\hbar$. Here the minimization means that we choose the minimal on-shell action if the classical pathes are not unique.

Let us consider the complexity between  the ground state $|\Omega\rangle$ of a Hamiltonian and the field operator eigenstate $|\varphi_0\rangle$.
The field operator eigenstate is the continuum limit of the product state in the configuration space and it is assumed to be the reference state in the path-integral complexity.
The inner product between these two states can be computed by the Euclidean path integral as follows:
\begin{equation}\label{innerOmega3}
\begin{split}
{Z_E := \langle\varphi_0(x)|\Omega\rangle }=\frac1{\mathcal{N}}\int_{\varphi(x,0)=\varphi_0(x)}\pD[\varphi(x)]\exp\left\{-\frac{1}{\hbar}S_E[\varphi]\right\}\,.
  \end{split}
\end{equation}
Here,  $\varphi_0(x) = \langle x | \varphi_0  \rangle$, $S_E[\varphi]$ is the Euclidean action, and {the normalization factor $\mathcal{N}$ is
\begin{equation}
\mathcal{N}:=\int_{\varphi(x,0)=\Omega(x)}\pD[\varphi(x)]\exp\left\{-\frac{1}{\hbar}S_E[\varphi]\right\}\,,
\end{equation}
with  $|\langle\Omega|\Omega\rangle|=1$.}

The absolute value symbol in Eq.~\eqref{innerOmega3} is not necessary because the function in the integration is positive. The upper bound of integration is omitted: in the Euclidean case, $\varphi(x,\infty)$ is the ground state $\Omega (x)= \langle x| \Omega \rangle$ so we do not need to specify it.
In the limit $\hbar\rightarrow0$, the complexity between $|\Omega\rangle$  and $|\varphi_0(x)\rangle$ is approximately~\footnote{In the Euclidean path integral, there is no imaginary part.}
\begin{equation}\label{innerOmega3b}
\begin{split}
 \bar{\C} &= {-\ln |\langle\varphi_0(x)|\Omega\rangle| =  -\ln Z_E }   \\
  &\approx \frac{1}{\hbar}\min\{S_{E, \text{on-shell}}[\varphi]-S_0\}\,,
 \end{split}
\end{equation}
where $S_{E, \text{on-shell}}[\varphi]$ is the Euclidean on-shell action and $S_0=\ln\mathcal{N}\approx S_E[\Omega]$ is the Euclidean on-shell action for the ground state ($|\phi_0\rangle = |\Omega \rangle$). Here the minimization means that we choose the minimal on-shell action if the classical pathes are not unique.

The equations~\eqref{Conshell1} and \eqref{innerOmega3b} show that our proposal \eqref{defholoCeq2} indeed define a kind of ``complexity'' if we use the action (or Euclidean action) to define the cost. However, this complexity has many essential differences compared with the circuit complexity: (1) it does not allow people to choose the ``gates'' and penalties artificially; (2) it may be negative; (3) it is unitary invariant.



We now prove that our proposal~\eqref{defholoCeq2} satisfies the triangle inequality not only for the infinitesimally close states but also for general states which have classical correspondences.
Here, we assume that the classical trajectory is stable so the on-shell action is locally minimal. Consider three states $\{|\varphi_i\rangle\}$ (i=1,2,3), which are the eigenstates of field operator $\varphi$ and correspond to the classical field configurations  $\varphi_i|_x=\varphi_i(x)$.  Then according to our formula~\eqref{Conshell1}, up to the leading order we have
\begin{align}\label{Conshell1b1}
 \bar{\C}(|\varphi_i\rangle,|\varphi_j\rangle)= \frac{1}{\hbar}\min\{|S_{\text{cl}}[\varphi]| \   | \ \forall\varphi(x,t), \ s.t. \ \varphi(x,t_1)=\varphi_i(x),~\varphi(x,t_2)=\varphi_j(x)\}\,,
\end{align}
with $i,j=1,2,3$. By this formula, we find
\begin{equation}\label{varphitrig1}
  \bar{\C}(|\varphi_i\rangle,|\varphi_j\rangle)+\bar{\C}(|\varphi_j\rangle,|\varphi_k\rangle)\geq \bar{\C}(|\varphi_i\rangle,|\varphi_k\rangle),~~i,j,k=1,2,3\,.
\end{equation}
Thus, for the quantum states which have classical correspondences, our formula  gives us a kind of ``distance''.

Let us make a few comments. Our proof of the triangle inequality Eq.~\eqref{varphitrig1} here is valid only for the quantum states which have classical correspondences. Otherwise, it is out of our scope.
For example, we do not mean to apply our proposal ~\eqref{defholoCeq2}  to some states which appear in quantum information processes, quantum circuits or quantum computations which do not have classical correspondences and even may not have Lagrangian formalism. In these cases, there are well-developed complexity theories in that context.

As we have repeatedly emphasized, the purpose of this paper is not to prove the holographic complexity is really equivalent to ``circuit complexity'' in every sense. It is still an open question and it is not something obvious a priori. We try to understand, without any prejudice, what the possible field theory dual of holographic complexity should be and then try to find certain relationships to the ``complexity'' of quantum circuits or quantum computations, if any.

Note that the boundary theory of an asymptotically AdS spacetime is not an arbitrary tunable quantum theory as in quantum circuits or quantum computations. Instead, it is a field theory which has classical correspondence. Our formula~\eqref{defholoCeq2} is designed for such cases and, for such cases, we find our proposal satisfies the triangle inequality.

From Eq.~\eqref{innerOmega3}, we can show that the ``path-integral complexity'' conjecture~\cite{Caputa:2017urj,Caputa:2017yrh} can be justified. In this ``path-integral complexity'', it was conjectured that complexity between certain states in two dimensional CFTs is given by the Liouville action.  There is a diagrammatic argument~\cite{Czech:2017ryf} why complexity is proportional to the Liouville action by using the relation between discretized path integrals and tensor network renormalization~\cite{PhysRevLett.115.180405}. The proof is  similar to what we have done in Ref.~\cite{Yang:2019udi} and we show it briefly in appendix~\ref{app1}. Here, our result is algebraic and the starting point has nothing to do with the tensor network renormalization. This agreement between different perspectives is a good supporting evidence for our proposal.


\subsection{Connections to holographic conjectures}
In this part we will show how the CV and CA conjectures can be understood from our proposal and explain what are the reference states in both conjectures. We also explain why two conjectures show different behaviors at early time~\cite{Carmi:2017jqz,Kim:2017qrq}. These will serve as supporting evidences for our proposal. Similar arguments have been shown in our previous work Ref.~\cite{Yang:2019udi}. We will show here that, for our more general proposal~\eqref{defholoCeq2}, we can still obtain the same conclusions.

\subsubsection{CV conjecture}
Let us start with a CFT Hamiltonian $H_0$ and its ground state $|\Omega\rangle$. Next, we consider a perturbation of the Hamiltonian by $H_I$: $H_\delta=H_0+ H_I\delta$, where $\delta$ is an infinitesimal parameter.  Then, we will have the perturbed ground state $|\Omega_\delta\rangle$. Denoting  the Euclidean Lagrangians of two Hamiltonians by $\mathcal{L}_0$ and $\mathcal{L}_\delta$, we have~\cite{MIyaji:2015mia,Alishahiha:2017cuk}
\begin{equation}\label{cvinner1}
  \langle\Omega_\delta|\Omega_0\rangle=\frac1{\sqrt{Z_0Z_\delta}}\int\pD\phi\exp\left[-\td^dx\left(\int_{-\infty}^0\td\tau\mathcal{L}_0 +\int_0^{\infty}\td\tau\mathcal{L}_\delta\right)\right] \,,
\end{equation}
where $\phi$ is the field variable. Eq.~\eqref{cvinner1} can be expanded as~\cite{MIyaji:2015mia,Alishahiha:2017cuk}
\begin{equation}\label{fi2Omega}
  \langle\Omega_\delta|\Omega\rangle=1-G_{\delta\delta}\delta^2+\mathcal{O}(\delta^4) \,,
\end{equation}
where the real value $G_{\delta\delta}$ is named fidelity of susceptibility~\cite{doi:10.1142/S0217979210056335} or the information metric~\cite{MIyaji:2015mia}. Thus, we find a simple relationship between the complexity and information metric at small $\delta$ limit
\begin{equation}\label{relC2IM}
  \bar{\C}(|\Omega\rangle,|\Omega_\delta\rangle)=-\ln|\langle\Omega_\delta|\Omega\rangle|=G_{\delta\delta}\delta^2\propto G_{\delta\delta} \,,
\end{equation}
Furthermore, it has been shown ~\cite{MIyaji:2015mia,Alishahiha:2017cuk} that, in conformal field theories  perturbed by a primary operator, the information metric is approximately a volume of the maximal time slice in the AdS spacetime, i.e.,
\begin{equation}\label{IMVs1}
  G_{\delta\delta}\propto\max_{\partial \Sigma=t_L\cup t_R}\text{Vol}(\Sigma)\,.
\end{equation}
Thus, by Eq.~\eqref{IMVs1} and Eq. \eqref{relC2IM}, we have
\begin{equation}\label{IMVs2}
 \bar{\C}(|\Omega\rangle,|\Omega_\delta\rangle)\propto\max_{\partial \Sigma=t_L\cup t_R}\text{Vol}(\Sigma) \,,
\end{equation}
which is nothing but the CV conjecture.

Note that the ground state of a CFT in holography is the TFD state dual to the double-sided black hole geometry. Thus, according to our proposal, the complexity in the CV conjecture may be interpreted as the complexity between the TFD state and its perturbed TFD state by a marginal operator. By this way, we clarified the reference state in the CV conjecture, while, in most literatures, it is just assumed to be an unknown ``simple'' reference state.

\subsubsection{CA conjecture}
Regarding the CA conjecture
we first consider Euclidean cases. By Eq.~\eqref{innerOmega3b} the complexity between the ground state and the  field operator eigenstate is obtained by the partition function of the boundary field theory
\begin{equation}\label{C2Zbd1}
  \bar{\C}=-\ln Z_{\text{bd}}[\phi(x)] \,.
\end{equation}
However, according to the AdS/CFT correspondence the partition function of the boundary field theory is dual to the one of a bulk gravity theory:
\begin{equation}\label{AdSCFTZ1}
  Z_{\text{bd}}[\phi(x)]=Z_{\text{bulk}}[g_{\mu\nu},\phi(x,z)]\,,
\end{equation}
with the matter fields satisfying the boundary condition $\phi(x,z)|_{z=0}=\phi(x)$. 
Thus, from Eq.~\eqref{C2Zbd1} and Eq.~\eqref{AdSCFTZ1} we have
\begin{equation}\label{C2Zbd2}
  \bar{\C}=-\ln \int\pD[g_{\mu\nu}]\pD[\phi]\exp\left\{-\frac1{\hbar}S_E[g_{\mu\nu}\,,\phi(x,z)]\right\}\,,
\end{equation}
where $S_E$ is the Euclidian action of the bulk gravity. In the weak gravity  limit,
\begin{equation}\label{C2Zbd3}
  \bar{\C}\approx\frac1{\hbar}S_{E,\text{on-shell}}[g_{\mu\nu},\phi(x,z)]=\frac1{\hbar}\left[\int_{I}\td t\int_{V(t)}H_E(g_{\mu\nu},\phi)\td^dx+S_{\text{bd}}\right]\,,
\end{equation}
where $H_E(g_{\mu\nu},\phi)$ is the Euclidean Hamiltonian density and $S_{\text{bd}}$ is a suitable boundary term. $V(t)$ is a time slice in the bulk at time $t$ and $I$ stands for the integration domain of Euclidean ``time'' $t$, both of which  depend on the physical system itself. If we consider the thermal system (including vacuum state), then $V(t)$ is a static time slice and $I=[0,\beta]$ with a periodic boundary condition at $t=0$ and $t=\beta$.


In the Lorentzian case, we consider the complexity between two ``field operators eigenstates''. The complexity is given by $\bar{\C}=-\ln|Z|+\text{Im}\ln Z$, where $Z$ is the inner product of two ``field operators eigenstates'' and is given by the path-integral
\begin{equation}
Z=\int\pD[g_{\mu\nu}]\pD[\phi]\exp\left\{\frac{i}{\hbar}S[g_{\mu\nu}\,,\phi(x,z)]\right\}\,.
\end{equation}
In the limit $\hbar\rightarrow0$, $\bar{\C}$ is dominated by the imaginary part of $Z$ and so we have
\begin{equation}\label{C2Zbd4}
  \bar{\C}\approx\frac1{\hbar}|S_{\text{on-shell}}[g_{\mu\nu},\phi(x,z)]|=\frac1{\hbar}\left|\int_{\mathcal{M}}\mathcal{L}(g_{\mu\nu},\phi)\td^{d+1}x+S_{\text{bd}}\right|\,,
\end{equation}
%
where $\mathcal{L}(g_{\mu\nu},\phi)$ is the Lagrangian density of the gravity theory.
To compute Eq.~\eqref{C2Zbd4},  the integration domain $\mathcal{M}$  needs to be clarified carefully.

\begin{figure}
  \centering
  \includegraphics[width=0.5\textwidth]{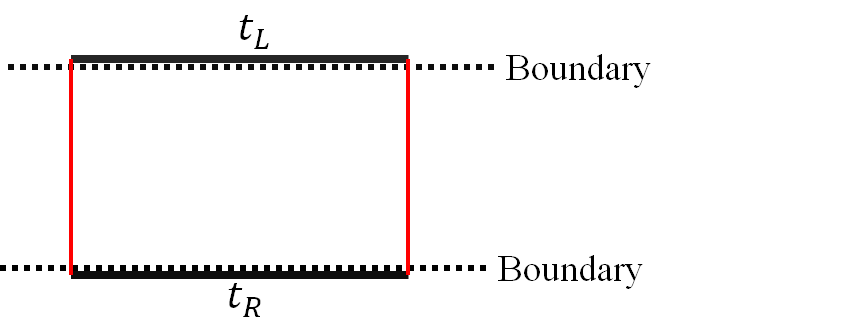}
  \includegraphics[width=0.38\textwidth]{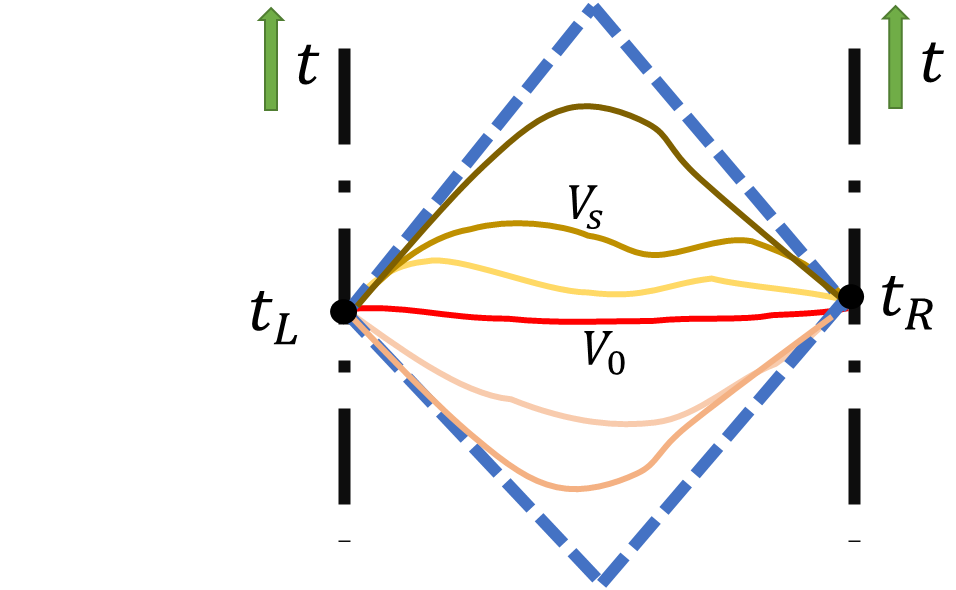}
  \caption{Left: In CA conjecture, the two disconnected time slices $t_L$ and $t_R$ are both infinitely large. We first assume that $t_L$ and $t_R$ are finite but large enough intervals. Right: in the Lorentzian case, when the two-side boundary are large enough, the entanglement wedge of $t_L\cup t_R$ is just its WdW patch. } \label{CAbulk}
\end{figure}

In Euclidean case, a boundary time slice stands for the target quantum states  so we have to choose bulk domain which is encoded into such time slice. In other word, we have to choose a bulk domain, of which all information could be reconstructed only by the boundary slice. Based on the AdS/CFT correspondence, we know such bulk region must be the entanglement wedge of the boundary slice. The entanglement wedge is defined in the full $d + 1$ dimensional spacetime as the causal domain of dependence of
the homology surface $V_0$, where $V_0$ is a $d$-dimensional surface which is surrounded by boundary slice and its corresponding Ryu-Takayanagi (RT) surface. In CA conjecture, the two disconnected time slices $t_L$ and $t_R$ are both infinitely large. However, physically it will be more convenient to first assume that $t_L$ and $t_R$ are finite but large enough intervals and take the infinite limit finally. For a pair of large enough two-side boundary slice $t_L\cup t_R$ in two-side black hole, the RT surface will connect these two disconnected time slices  so $V_0$ is a codimension-one  space-like surface which is attached at $t_L$ and $t_R$, see Fig.~\ref{CAbulk}. It is clear that the entanglement wedge in this case is just the WdW patch
\begin{equation}\label{boundaryCD}
  \mathcal{M}=\bigcup_{\partial V_s=t_L\cup t_R}V_s\,.
\end{equation}
where ${V_s}$ stands for an arbitrary space-like codimension-one surface connecting two-side boundary slice $t_L\cup t_R$.
Thus we obtain following result
\begin{equation}\label{C2Zbd5}
  \bar{\C}\approx\frac1{\hbar}\left[\int_{\text{WdW}}\mathcal{L}(g_{\mu\nu},\phi)\td^{d+1}x+S_{\text{bd}}\right]\,.
\end{equation}
The absolute sign disappears because it has been shown the on-shell action of WdW-patch is always positive~\cite{Carmi:2016wjl}. Eq.~\eqref{C2Zbd5} is nothing but the CA conjecture!

We find that the CA conjecture describes the complexity between the field operator eigenstate of a boundary field theory and a TFD state dual to the double-sided AdS black hole, in the holographic context. It is not the complexity between two TFD states.
The difference of the reference states for the CV and CA conjecture explains why they show different time-evolution  at early time, reported in~\cite{Carmi:2017jqz,Kim:2017qrq}.

\section{Applications to the TFD states}\label{tfds1}
In this section we will use our proposal to study the complexity of the TFD states. We will show our proposal can reproduce some results of the CA conjecture: both the divergent term and the finite term.

\subsection{Complexity of time-independent TFD states}
To compare with the CA conjecture, we consider a general TFD state
\begin{equation}\label{timeTFDstate00}
  |\text{TFD}\rangle:=\frac1{\sqrt{Z(\beta)}}\sum_{E_n}e^{-\beta E_n/2}|E_n\rangle_R|E_n\rangle_L\,,
\end{equation}
with the inverse temperature $\beta$, eigen-energy $E_n$ and partition function
\begin{equation}\label{defZbeta}
Z(\beta):=\sum_{E_n}e^{-\beta E_n}\,.
\end{equation}
As we have argued, in the CA conjecture, the reference state should be the eigenstate of the field operator.  To write down such a state, in general we need the detailed action of a theory. Thus, we cannot directly construct such a reference state only based on the general TFD state~\eqref{timeTFDstate00}. Nevertheless, this reference state is a kind of ``simple'' state which has no spatial entanglement and there is another type of ``no-spatial-entanglement'' state. We know that the strong thermal fluctuation will destroy any quantum correlation, so the TFD state at infinite temperature has no spatial entanglement. Based on this property, we use the following state as a reference state
\begin{equation}\label{refRT1}
  |\tilde{R}^{(\varepsilon)}_{\alpha}\rangle:=\frac1{\sqrt{Z(\varepsilon\alpha)}}\sum_{E_n}e^{-\varepsilon \alpha E_n/2}|E_n\rangle_R|E_n\rangle_L\,,
\end{equation}
where $\varepsilon\rightarrow0$ stands for the UV cut-off and $\alpha$ is an arbitrary positive constant. Though this state may not be an eigenstate of a field operator, it is still a kind of classical state with no spatial entanglement and we will show that it can be used as a reference state to reproduce basic properties of the CA conjecture.

As $\langle\tilde{R}^{(\varepsilon)}_{\alpha}|\text{TFD}\rangle$ is a real number, the complexity between $|\tilde{R}^{(\varepsilon)}_\alpha\rangle$ and $|\text{TFD}\rangle$ in our proposal reads
\begin{equation}\label{comptfd1}
  \bar{\C}=-\ln\langle\tilde{R}^{(\varepsilon)}_{\alpha}|\text{TFD}\rangle\,.
\end{equation}
It is easy to find
\begin{equation}\label{innerTFDs1}
  \langle\tilde{R}^{(\varepsilon)}_{\alpha}|\text{TFD}\rangle=\frac1{\sqrt{Z(\beta)Z(\alpha\varepsilon)}}\sum_{E_n}e^{-(\beta+\alpha\varepsilon) E_n/2}=\frac{Z(\beta/2)}{\sqrt{Z(\beta)Z(\alpha\varepsilon)}}\,,
\end{equation}
where we neglected the $\varepsilon$ term in $\beta+\alpha\varepsilon$. Then our proposal \eqref{comptfd1} gives
\begin{equation}\label{cforrtfd1}
  \bar{\C}=-\ln\frac{Z(\beta/2)}{\sqrt{Z(\beta)Z(\alpha\varepsilon)}}\,.
\end{equation}

For a CFT theory in $d$-dimensional spacetime, the partition function of a thermal field has the universal form
\begin{equation}\label{freeecft1}
  Z(\beta)\propto\exp(b_dV\beta^{1-d})=\exp(b_dVT^{d-1})\,,
\end{equation}
where $b_d$ is a constant and proportional to the central charge of the CFT. Plugging Eq.~\eqref{freeecft1} in Eq.~\eqref{cforrtfd1}  we obtain
\begin{equation}\label{cforrtfd2}
  \bar{\C}=\frac12\alpha^{1-d}b_dV[\varepsilon^{1-d}+\alpha^{d-1}(1-2^{2-d})T^{d-1}]\,.
\end{equation}
Note that the overall factor of complexity is irrelevant. It has been shown that the complexity of the CA conjecture in the AdS-Schwarzschild black hole~\cite{Chapman:2016hwi} is given by
\begin{equation}\label{cascheq1}
  \mathcal{C}=\frac{\ln d}{4\pi}V\varepsilon^{1-d}+\frac{d-2}{4\pi d}\cot(\pi/d)\left(\frac{4\pi}{d}\right)^{d-1}VT^{d-1}\,,
\end{equation}
where we set $G_N=1$. By choosing the parameter $\alpha$ suitably in \eqref{cforrtfd2}, we can recover the holographic result~\eqref{cascheq1} exactly.

It is interesting to compare our result with other proposals, which start from the complexity of quantum circuits: for example, the proposal of Ref.~\cite{Molina-Vilaplana:2018sfn} based on the cMERA or the proposal of Ref.~\cite{Chapman:2017rqy} based the Fubini-Study metric.
Both methods can reproduce the leading divergent term of the holographic complexity. However, In Ref.~\cite{Molina-Vilaplana:2018sfn}, $L^1$ norm is assumed artificially and, in Ref.~\cite{Chapman:2017rqy} the generators set is restrict to be SU(1,1) and $L^1$ norm is also assumed artificially. In addition, though the leading divergent term was reproduced in their work, the temperature-dependent term of Eq.~\eqref{cascheq1} cannot be reproduced.

In our framework, Eq.~\eqref{cforrtfd2} naturally appears without artificial assumptions such as $L^1$ norm or specific generators set.  We stress that we obtained the correct temperature-dependent term  as well as the divergence structure.
To our knowledge, this is the only case that a field theory proposal reproduce the complete form of the holographic result~\eqref{cascheq1}.

\subsection{Time evolution and compatibility with holographic results}
We also would like to make a short comment regarding the inner product between a TFD state and its time-evolution state at large time limit. A time-dependent TFD state is given by
\begin{equation}\label{timeTFDstate0066}
  |\text{TFD}(t)\rangle:=\frac1{\sqrt{Z(\beta)}}\sum_{E_n}e^{-(\beta+2it)E_n/2}|E_n\rangle_R|E_n\rangle_L\,.
\end{equation}
The complexity between $|\text{TFD}(t)\rangle$ and $|\text{TFD}(0)\rangle$ in our proposal reads
\begin{equation}\label{comptfd166}
  \bar{\C}(t)=-\ln|F(t)|+\text{Im}\ln F(t) \,,
\end{equation}
where
\begin{equation}\label{defFts100}
  F(t):=\langle\text{TFD}(t)|\text{TFD}(0)\rangle\,.
\end{equation}

In the CA conjecture, the reference state is conjectured to be a kind of ``simple'' state. In this paper, we have argued that this ``simple'' state should be an eigenstate of the field operator and we denote it by $|R\rangle$.  Though, in the previous subsection,  we used $|\tilde{R}^{(\varepsilon)}_\alpha\rangle$ \eqref{refRT1} as a reference state and reproduced a few basic properties of complexity in the CA conjecture, here we do not assume $|R\rangle=|\tilde{R}^{(\varepsilon)}_\alpha\rangle$.

Let us introduce three different complexities: i) $\bar{\C}_R(0)$ is the complexity between $|R\rangle$ and the initial TFD state $|\text{TFD}(0)\rangle$, ii) $\bar{\C}_R(t)$ is the complexity between $|R\rangle$ and the time-dependent TFD state  $|\text{TFD}(t)\rangle$, iii) $\bar{\C}(t)$ is the complexity between $|\text{TFD}(0)\rangle$ and $|\text{TFD}(t)\rangle$.
If the complexity in the CA conjecture stands for a kind of distance~\cite{Susskind:2014jwa,Brown:2016wib,Brown:2017jil}, the triangle inequality implies
\begin{equation}\label{triineq1}
  \bar{\C}_R(t)-\bar{\C}_R(0)\leq\bar{\C}(t)\,.
\end{equation}

In the CA conjecture, it has been discovered that the complexity between $|R\rangle$ and  $|\text{TFD}(t)\rangle$ will grow linearly at late-time limit. Thus, if we accept the result of the CA conjecture as $\bar{\C}_R(t)$, we may conclude that $\bar{\C}(t)$ should increase forever as time goes on: $\bar{\C}(\infty)\rightarrow\infty$, i.e. the complexity between between $|\text{TFD}(0)\rangle$ and $|\text{TFD}(t)\rangle$ should grow to infinity as $t\rightarrow\infty$. Let us check it concretely.

Plugging Eq.~\eqref{timeTFDstate0066} into Eq.~\eqref{defFts100} we obtain
\begin{equation}\label{innerTFDs1}
  F(t)=\frac1{Z(\beta)}\sum_{E_n}e^{-(\beta+it) E_n}=\frac{Z(\beta+it)}{Z(\beta)}\,.
\end{equation}
In Eq.~\eqref{innerTFDs1} we have
\begin{equation}
Z(\beta+it)=\sum_{n}e^{-\beta E_n}e^{iE_nt}\,.
\end{equation}
In the continuum limit, we may replace the sum with the integral:
\begin{equation}\label{intforfTFD}
  Z(\beta+it)=\int_0^\infty N(E)e^{-\beta E}e^{iEt}\td E\,,
\end{equation}
where the density of state $N(E)$ is introduced and $N(E)\td E$ is the state number when energy is in $E\sim E+\td E$. It is clear that
\begin{equation}\label{intforfTFD}
  \int_0^\infty \left|N(E)e^{-\beta E}\right|\td E=\int_0^\infty N(E)e^{-\beta E}\td E=Z(\beta)\,,
\end{equation}
which is finite (here we assume that volume $V$ is large but finite).
Then the Riemann-Lebesgue lemma says that
\begin{equation}\label{CFTZbetat1}
  \lim_{t\rightarrow\infty}Z(\beta+it)=\lim_{t\rightarrow\infty}\int_0^\infty N(E)e^{-\beta E}e^{-itE}\td E=0\,,
\end{equation}
so
\begin{equation}
\lim_{t\rightarrow\infty} F(t)=0\,.
\end{equation}
It implies that the complexity between $|\text{TFD}(t)\rangle$ and $|\text{TFD}(0)\rangle$ will grow forever.\footnote{See Ref.~\cite{Hashimoto:2018bmb} for another way of computation of $F(t)$ by an analytic continuation of Eq.~\eqref{freeecft1}. In our opinion, it seems that a simple analytical continuation may be misleading.}
This is another nice consistency check of our proposal with the CV conjecture.  As we do not know the reference state exactly just from the general `formal definition'' of the TFD state \eqref{timeTFDstate00}, we cannot directly compute $\bar{\C}_R(t)$. Thus, the consistency check we just showed is the best we can do.

\section{Conclusions}\label{summ}
Complexity is a quantum informational quantity, which essentially depends on the choices of the basic operations (gates) and their costs (penalties).  There are proposals for the holographic duals of the complexity.
In the complexity-volume and the complexity-action conjectures, the complexity is conjectured to be the volume of a maximal spatial hypersurface or the on-shell action of the bulk theory in a special spacetime region. In both conjectures, they neither tell us what the fundamental operators (gates) are nor tell us how to choose the costs (penalties).
Thus, there are two fundamental questions: if the holographic complexity is indeed a kind of complexity, in their field theory duals, what are the basic gates and their costs?

Towards the answer to these fundamental questions, we choose a different strategy. We do not assume anything from usual concepts from the complexity in quantum information theory. In particular, we do not assume that the field theory dual of holographic complexity is a continuum version of the circuit complexity. Without any prejudice, we start with the inherent properties of the holographic complexity and try to understand the essential features that the field theory duals should have.  In principle, the dual may not be any kind of complexity.

We argue that any field theory dual of the holographic complexity should have two basic properties:  (1) it is invariant under infinitely many independent unitary transformations and (2) it is extensive for product states. These two basic properties are inferred from the holographic complexity itself but cannot be obtained by the general analyses of quantum informational setups including ``circuit complexity'' or ``operators complexity''.
Guided by these two properties, we proposed a possible candidate for the field theory dual of the holographic complexity:
\begin{equation}\label{defholoCeq2b}
  \bar{\C}(|\psi_1\rangle,|\psi_2\rangle)=\alpha_1\ln|\langle\psi_1|\psi_2\rangle|+\alpha_2\left|\text{Im}\ln\langle\psi_1|\psi_2\rangle\right|\,.
\end{equation}
%
This simple-looking formula has rich contents.

Firstly, the complexity in field theory can have a natural path integral formalism. Though formula~\eqref{defholoCeq2b} does not satisfy the ``triangle inequality'' in usual qubit systems, we show that it can satisfy the triangle inequality for a theory which has the Lagrangian formalism. In addition, in the classical limit, it naturally gives the relation between the ``path-integral complexity'' and the Liouville action for 2D conformal field theories.

Secondly, the proposal~\eqref{defholoCeq2b} can give natural interpretations for the CV and CA conjectures and clarified their target and reference states. The CV conjecture is dual to the complexity between a TFD state and its perturbed state by a marginal operator. The CA conjecture computes the complexity between a TFD state and the eigenstate of the field operator. This difference explains why the two holographic conjectures have different time evolution.

Finally, if our proposal is applied to the TFD states we find that it is compatible with the holographic complexities at late time. Furthermore, our proposal can reproduce the holographic results of the CA conjecture well. It yields both the correct divergent structure  and the temperature-dependent term. To our knowledge, this is the only field theory proposal that reproduces the both terms in the result from the CA conjecture.

One basic property of our proposal~\eqref{defholoCeq2b} is that it is unitary invariant. It is often claimed~\cite{Brown:2017jil, Balasubramanian:2019wgd} that the complexity must be non-unitary invariant because a unitary-invariant complexity cannot reproduce the ``expected'' time evolution of the complexity: for a chaotic system with $N$ degrees of freedom, the complexity evolves in three stages: linear growth until $t\sim e^{N}$, saturation and small fluctuations, and quantum recurrence at $t\sim e^{e^N}$.
However, the counter example of this claim is shown in Ref.~\cite{Yang:2019iav}, where the unitary-invariant or bi-invariant complexity can indeed realize the expected time evolution. The example in \cite{Yang:2019iav} is a supporting evidence for our claim that the field theory dual of the holographic complexity may be unitary-invariant.  What is more, the unitary invariance also matches with the fact that holographic complexity is diffeomorphic invariant. Every bulk diffeomorphic transformation will induce a unitary transformation on the boundary states, so the boundary complexity should be invariant under infinitely many independent unitary transformations. To the best of our knowledge, there is no non-unitary invariant complexity which is invariant under these infinitely many independent unitary transformations.

We want to emphasize that there is nothing wrong with the ``non-unitary-invariance'' of the complexity in {\it real quantum circuits}. The essential question we are asking in this paper is ``what is the boundary field theory dual corresponding to the holographic complexity?''. Therefore, the properties of the complexity of the real circuits are never requirements. They must be consequences, if possible. If we assume that the holographic complexity is dual to a kind of continuum version of the discrete circuit complexity, in our opinion, some of the properties of the real circuit complexity need to be modified to satisfy two properties we proposed of this paper.

One may argue that, our proposal is basically a certain function of an inner product, which describes some properties of ``overlap'' between two states. Particularly, it has been noted that the function of a inner product, such as the Fubini-Study distance, can not distinguish one-flip from multi-flips~\cite{Brown:2017jil} of qubits systems, so may not be a good candidate of complexity in qubits systems. Regarding this viewpoint, we want to make three comments. First, we have shown in the end of Secs.~\ref{proptrig} and \ref{trigineq} that our proposal satisfies the triangle inequality so stands for a kind of ``distance''.
Second, we have shown in the appendix~\ref{answers1} how our proposal can distinguish one-flip from multi-flips if it is used for ``continuous systems'' rather than discrete qubits systems. Thirdly, if a simple function of the inner product can reproduce most of basic properties of holographic complexity, there is no reason to naively abandon the possibility that holographic complexity in fact describes some properties of ``overlap'' of boundary states. Holographic complexity may or may not be related to kind of naively generalization of circuit complexity, which is still an open question, to our understanding.
We believe that looking at the problem from a different angle will be a meaningful starting point to understand holographic complexity better.


\acknowledgments
The work of K.-Y. Kim was supported by Basic Science Research Program through the National Research Foundation of Korea (NRF) funded by the Ministry of Science, ICT $\&$ Future Planning(NRF- 2017R1A2B4004810) and GIST Research Institute(GRI) grant funded by the GIST in 2021. C. Niu is supported by the Natural Science Foundation of China under Grant No. 11805083. C.Y. Zhang is supported by Project funded by China Postdoctoral Science Foundation. R.-Q. Yang is supported by the Natural Science Foundation of China under Grant No. 12005155.

\appendix
\section{Proof for path-integral complexity}\label{app1}
For a two-dimensional conformal field theory which contains matter fields coupling with string worldsheet and is embedded in a D-dimensional flat space ($2<D<25$), the classical action is
\begin{equation}\label{CFTworldsheet}
  S:=(2\pi\alpha)^{-1}S_X+S_m[\varphi,g_{ab}]\,.
\end{equation}
Here, $S_X:=\int\td^2x g^{ab}\partial_aX^\mu\partial_bX^\nu\eta_{\mu\nu}$ stands for the string worldsheet action with
 the Minkowski metric $\eta_{\mu\nu}$ in the D-dimensional background space. $g_{ab}$ is the induced metric of the worldsheet.  $\alpha$ is the string coupling constant, which is proportional to string length square. $S_m$ is a conformal matter fields action.

The Euclidian action reads~\cite{POLYAKOV1981207,DAS1989,Ginsparg:1993is}
\begin{equation}\label{CFTworldsheet}
  S_E =S_m[\varphi,\delta_{ab}]+\frac{1}{2\pi\alpha}(S_X[X^\mu,\delta_{ab}]+S_L[\phi,\delta_{ab}]+S_{gh}[b^{ab},c_a,\delta_{ab}])\,,
\end{equation}
where  $S_{gh}$ is the ghost fields action and $S_L$ is the Liouville action with the  central charge $c$:
\begin{equation}
S_L[\phi,\delta_{ab}]:={c}/(24\pi)\iint\td^2 x\left[\eta^{ab}\partial_a\phi\partial_b\phi+\mu e^{2\phi}\right] \,.
\end{equation}
Suppose that a common eigenstate of $\{\varphi=\varphi_0, X^\mu=X^\mu_0, g_{ab}^{(E)}=\delta_{ab}\}$ is $|\varphi_0\rangle$; and $|\Omega_\phi\rangle$ stands for the ground state satisfying $g_{ab}^{(E)}|_{z=z_0}=e^{2\phi(x)}\delta_{ab}$, where $z$ is the Euclidean time and $z_0=\epsilon\ll1$ is a UV cut-off.
Then we have
\begin{equation}\label{Liuo1s}
\begin{split}
  \langle\varphi_0|\Omega_\phi\rangle=&\int\pD[\phi]\pD[\varphi]\pD[X]\pD[b]\pD[c]\exp\left\{-\frac{1}{\hbar}S_E\right\}\\
  =&\left[\int\pD[\phi]\exp\left(-\frac{S_L}{2\pi\alpha\hbar}\right)\right]\langle\varphi_0|\Omega_0\rangle\,.
  \end{split}
\end{equation}
Here $|\Omega_0\rangle$ stands for the ground state for $\phi=0$. Based on our proposal~\eqref{defholoCeq2} and noting the fact that there is no imaginary part, we find that the complexity between $|\varphi_0\rangle$ and $|\Omega_\phi\rangle$ reads
\begin{equation}\label{Cmforphi1}
  \bar{\C}[\phi]=-\ln\int_{\phi(x,z=\epsilon)=\phi(x)}\pD[\phi]\exp\left(-\frac{S_L}{2\pi\alpha\hbar}\right) -\ln\langle\varphi_0|\Omega_0\rangle\,,
\end{equation}

Let us compare our result Eq.~\eqref{Cmforphi1} with the path integral complexity in Refs.~\cite{Caputa:2017urj,Caputa:2017yrh}.
In the small $\hbar \alpha$ limit, Eq.~\eqref{Cmforphi1} yields, by the saddle point approximation,
\begin{equation}\label{Cmforliovs2}
  \bar{\C}=\bar{\C}(0)+\frac{S_L^{(cl)}[\phi]}{2\pi\hbar\alpha}[1+\mathcal{O}(\hbar\alpha)]\,,
\end{equation}
where $S_L^{(cl)}[\phi]$ is the classical on-shell action of the Liouville action with the boundary condition $\phi(x,\epsilon)=\phi(x)$ \footnote{There are two different limits that we can recover the proposal for the Liouville action: $\hbar\rightarrow0$ and $\alpha\rightarrow0$. The former is the usual classical limit while the later is the weak coupling limit between the matter and string/gravity.} and $\bar{\C}(0)$ corresponds to $S_0$ in Eq \eqref{innerOmega3b}, which contains all terms which are independent of the dynamics of matters. Thus, we find that the conjecture of Refs.~\cite{Caputa:2017urj,Caputa:2017yrh} is just the leading order term of our proposal in the classical limit.

\section{A comment on Fubini-Study distance}\label{answers1}
In this appendix, we will show that how our proposal can distinguish one-flip from multi-flips if it is used for  ``continuous systems'' rather than discrete qubits systems. The basic idea of argument was shown in our previous work~\cite{Yang:2019udi}. We explain here again for the readers.

To understand our idea, let us first consider the Fubini-Study distance, which was discussed by Ref.~\cite{Brown:2017jil} and used to explain a function of inner product cannot be used to compute the complexity in qubit systems. The Fubini-Study distance between two product states $|\psi_1\rangle=\bigotimes_{i=1}^{n}|a_n\rangle$ and $|\psi_2\rangle=\bigotimes_{i=1}^{n}|b_n\rangle$ is given by
\begin{equation}\label{FBdist2}
  D_{FS}=\arccos\left|\prod_{i=1}^n\langle a_i|b_i\rangle\right|\,,
\end{equation}
Here we take $a_i=0,1$ and $b_i=0,1$ so that $|\psi_1\rangle$ and $|\psi_2\rangle$ are two quibit states. Let us consider $|\psi_1\rangle = |\psi_2 \rangle$ as a starting point, i.e. two states are same states and so the complexity is zero. Only one flip of a qubit will change the complexity dramatically: from {zero} to $\pi/2$. This can not be the property of the complexity because flipping just one qubit should not change the complexity that much. If we start with the case $\langle \psi_1 | \psi_2 \rangle = 0$ flipping some of the qubits may not change the complexity at all. This can not be the property of the complexity either.

However, this this does not mean that we can not use the inner product at all for complexity. We will show that the above issue may be resolved by our proposal~\eqref{defholoCeq2}:
\begin{equation}\label{nqubitsum}
  \bar{\C}=\bar{\C}_r+\bar{\C}_{\text{Im}}\,,
\end{equation}
where
\begin{equation}\label{nqubitsum}
  \bar{\C}_r=-\sum_{i=1}^n\ln|\langle a_i| b_i\rangle|\,, \quad \bar{\C}_{\text{Im}}=\left|\text{Im}\sum_{i=1}^n\ln\langle a_i| b_i\rangle\right|\,.
\end{equation}
Though here the function ``ln'' replaces the function ``arccos'', the problem of ``flip one qubit'' is still seemingly unsolved. We will argue that  this issue can be resolved if we take into account that our formula~\eqref{defholoCeq2} is proposed for the system in the \textit{continum limit with infinitely many degrees of freedoms} rather than discrete finite systems.


Let us focus on the real part $\bar{\C}_r$.
For a 1 dimensional a continuous system
\begin{equation}\label{contqubit11}
\bar{\C}_r=-\ln|\langle\psi_1|\psi_2\rangle|=-2L\int\ln|\langle a(k)|b(k)\rangle|\td k\,,
\end{equation}
where  the continuous variable $k$ is introduced instead of the discrete index $i$.  $L$ has the dimension of $[k]^{-1}$.
Suppose that the states are `regular', meaning that the inner product $\langle a(k)|b(k)$ depends on $k$ analytically. If we change the states of $k\in(k_0-\delta,k_0+\delta)$ so that $\langle a(k_0)|b(k_0)\rangle\neq0\rightarrow\langle \tilde{a}(k_0)|\tilde{b}(k_0)\rangle=0$, the complexity is changed as
\begin{equation}\label{contqubit1b}
\delta\bar{\C}_r=-L\int_{-\delta}^{\delta}[\ln|\langle \tilde{a}(k_0+x)|\tilde{b}(k_0+x)\rangle|-\ln|\langle a(k_0+x)|b(k_0+x)\rangle|]\td x\,,
\end{equation}
and the integral is finite due to `log', although  $\langle \tilde{a}(k_0+x)|\tilde{b}(k_0+x)\rangle$ is zero at $x=0$. In the limit $\delta\rightarrow0$, which amounts to ``flipping exactly one qubit'',  $\delta\bar{\C}_r=0$ as expected. The same holds also for $\bar{\C}_{\text{Im}}$.
%

We can use our proposal for the discrete system by making a discretization on the integration measure ``$\int\td k$'' and a regularization in the argument of ``$\ln$''. For a $n$-qubit system, one way is
\begin{equation}
L\int\td k\rightarrow\sum \,,
\end{equation}
for a discretization and
\begin{equation} \label{coff123}
\ln\langle\cdot|\cdot\rangle\rightarrow\ln(\langle\cdot|\cdot\rangle+\bar{\varepsilon}) \,,
\end{equation}
with $\bar{\varepsilon}\ll1$ for a regularization. The discrete version of Eq.~\eqref{contqubit11} is
\begin{equation}\label{contqubit111}
  \bar{\C}_r=-\sum_{i=1}^n\ln\left(\frac{|\langle a_i|b_i\rangle|+\bar{\varepsilon}}{1+\bar{\varepsilon}}\right),
\end{equation}
where $1+\bar{\varepsilon}$ in the denominator is introduced to ensure that the complexity between the same states is zero. Let us start with the case $|\psi_1\rangle = |\psi_2 \rangle$, which yields $\bar{\C}(|\psi_2\rangle,|\psi_1\rangle) =0$. Once we flip one qubit in $|\psi_2 \rangle$, the complexity increased by $\bar{\C}_0 = \bar{\C}_r=-2\ln(\bar{\varepsilon}/(1+\bar{\varepsilon})) \sim -2\ln \bar{\varepsilon} $. If we flip $n$ qubits in $|\psi_2 \rangle$ the complexity is increased by $n\bar{\C}_0$. It is a desirable property of the complexity.

The cut-off term $\bar{\varepsilon}$ in Eq.~\eqref{coff123} may look artificial. However, this can be understood in the following way. If we want to make two qubits $|a_{\text{th}}\rangle$ and $|b_{\text{th}}\rangle$ 
 it is easy to write down mathematically, but physically we have to design certain physical systems to realize them. Because of ubiquitous quantum and thermal fluctuations, what we really deal with are two states $|a_{\text{ob}}\rangle$ and $|b_{\text{ob}}\rangle$.  Their inner product is
\begin{equation}
|\langle a_{\text{ob}}|b_{\text{ob}}\rangle|=|\langle a_{\text{th}}|b_{\text{th}}\rangle|+\varepsilon\,,
\end{equation}
so
\begin{equation}
\bar{\varepsilon}=\overline{|\langle a_{\text{ob}}|b_{\text{ob}}\rangle|-|\langle a_{\text{th}}|b_{\text{th}}\rangle|} \,,
\end{equation}
where the ``$\overline{X}$'' means the average observations of the variable $X$ and $\bar{\varepsilon}$ is an ``error'' due to the intrinsic fluctuations. If $|\langle a_{\text{th}}|b_{\text{th}}\rangle|>0$,  $|\langle a_{\text{ob}}|b_{\text{ob}}\rangle|$ can be bigger or smaller than $|\langle a_{\text{th}}|b_{\text{th}}\rangle|$ , i.e. $\varepsilon$ can be positive or negative, so $\bar{\varepsilon}\rightarrow0$. However, if $|\langle a_{\text{th}}|b_{\text{th}}\rangle|=0$, we have $|\langle a_{\text{ob}}|b_{\text{ob}}\rangle|\geq0$ so $\bar{\varepsilon}>0$. The value of $\bar{\varepsilon}>0$ is determined by the physical limit of observation. In principle we can reduce it by considering complicated systems, which will increase the complexity as expected. It is reflected in our formula $\bar{\C}_0  \sim -\ln \bar{\varepsilon} $ in the previous paragraph.

\bibliographystyle{JHEP}
\bibliography{FG-ref-state}

\end{document}